\documentclass{PoS}\title{Pion Elastic and $\pi\to\gamma\,\gamma^*$
Transition Form Factors at Large Momentum
Transfers}\ShortTitle{Pion Elastic and $\pi\to\gamma\,\gamma^*$
Transition Form Factors at Large Momentum Transfers}\author{Irina
Balakireva\\D.~V.~Skobeltsyn Institute of Nuclear Physics, Moscow
State University, 119991, Moscow, Russia\\E-mail:
\email{iraxff@mail.ru}}\author{Wolfgang Lucha\\Institute for High
Energy Physics, Austrian Academy of Sciences, Nikolsdorfergasse
18, A-1050 Vienna, Austria\\E-mail:
\email{Wolfgang.Lucha@oeaw.ac.at}}\author{\speaker{Dmitri
Melikhov}\\Institute for High Energy Physics, Austrian Academy of
Sciences, Nikolsdorfergasse 18, A-1050 Vienna, Austria,\\Faculty
of Physics, University of Vienna, Boltzmanngasse 5, A-1090 Vienna,
Austria, and\\D.~V.~Skobeltsyn Institute of Nuclear Physics,
Moscow State University, 119991, Moscow, Russia\\E-mail:
\email{dmitri\_melikhov@gmx.de}}

\abstract{We present the results of our recent analyses of the
form factors $F_\pi(Q^2)$ and $F_{P\gamma}(Q^2)$,
$P=\pi,\eta,\eta',$ within the local-duality (LD) version of QCD
sum rules \cite{blm2011,lm2011}. To probe the expected accuracy~of
this method, we consider, in parallel to QCD, a quantum-mechanical
(QM) potential model. In~the latter case, the exact form factor
may be calculated from the solutions of the Schr\"odinger equation
and confronted with the result from the QM LD sum rule. We find
that the LD sum rule is expected to yield reliable predictions for
both $F_\pi(Q^2)$ and $F_{\pi\gamma}(Q^2)$ in the region
$Q^2\ge5$--$6$ GeV$^2$. Moreover, in this region the accuracy of
this approach improves rather fast with increasing $Q^2.$ For the
elastic form factor $F_\pi(Q^2),$ we are therefore forced to
conclude that large deviations from the LD limit in the region
$Q^2=20$--$50$ GeV$^2$ reported in some recent theoretical studies
seem to us unlikely. The data on the
$\eta,\eta'\to\gamma\,\gamma^*$ transition form factors meet
pretty well the predictions of an ``LD model.'' Interestingly,
recent {\sc BaBar} results for the $\pi^0\to\gamma\,\gamma^*$
transition form factor hint at an LD violation rising with $Q^2;$
this is at odds with the $\eta,\eta'$ cases and all our
experience~from~quantum~mechanics.}

\FullConference{The 2011 Europhysics Conference on High Energy
Physics, EPS-HEP 2011,\\July 21--27, 2011\\Grenoble,
Rh\^one-Alpes, France}

\begin{document}\section{Introduction}
\begin{figure}[!h]\begin{center}\begin{tabular}{c}
\includegraphics[width=7.89cm]{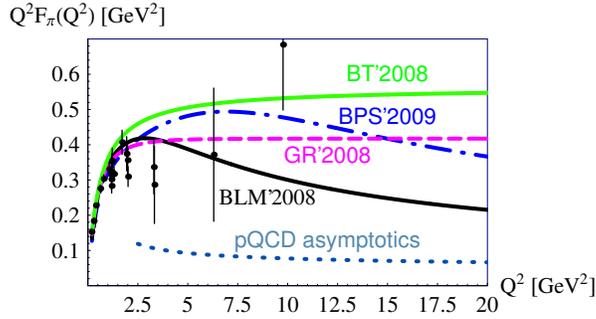}
\end{tabular}\caption{\label{Plot:1}Pion elastic form factor
$F_\pi(Q^2)$: recent theoretical predictions \cite{blm2011,recent}
vs.\ experimental~data~\cite{data_piff}.}\end{center}\end{figure}

The pion is full of
surprises: In spite of the long history of theoretical studies of
the pion~elastic form factor, no consensus on its behaviour in the
region $Q^2\approx5$--$50$ GeV$^2$ has been reached
(Fig.~\ref{Plot:1}); recent {\sc BaBar} results on the
$\pi\to\gamma\,\gamma^*$ form factor \cite{babar} imply a large
violation of pQCD factorization in a range of $Q^2$ up to 40
GeV$^2$. In \cite{blm2011,lm2011}, we investigated $F_\pi(Q^2)$
and $F_{P\gamma}(Q^2)$ by local-duality~(LD) QCD sum rules
\cite{ld}; their attractive feature is to offer the possibility to
study form factors of hadrons without knowing subtle details of
their structure and to consider different hadrons on equal
footing.

\section{Local-Duality Sum Rules in QCD}LD sum rules are
dispersive sum rules in the limit of infinite Borel mass
parameter: all power corrections vanish and all details of
nonperturbative dynamics are subsumed in a single quantity, the
{\em effective threshold\/} $s_{\rm eff}(Q^2)$. The basic objects
for finding form factors are three-point functions: for the pion
elastic form factor the $\langle AVA\rangle$ correlator, for the
transition form factor the $\langle AVV\rangle$ correlator, with
$A$ the axialvector and $V$ the vector current. Upon implementing
standard quark--hadron duality, sum rules relate these pion form
factors to the low-energy portions of the perturbative
contributions:\begin{equation}\label{Fld}
F_{\pi}(Q^2)=\frac{1}{f_\pi^2}\int\limits_0^{s_{\rm
eff}(Q^2)}\hspace{-1.5ex}{\rm d}s_1\int\limits_0^{s_{\rm
eff}(Q^2)}\hspace{-1.5ex}{\rm d}s_2\,\Delta^{(AVA)}_{\rm
pert}(s_1,s_2,Q^2)\ ,\qquad
F_{\pi\gamma}(Q^2)=\frac{1}{f_\pi}\int\limits_0^{\bar s_{\rm
eff}(Q^2)}\hspace{-1.5ex}{\rm d}s\,\sigma^{(AVV)}_{\rm
pert}(s,Q^2)\ ,\hspace{2ex}\end{equation}with double and single
spectral densities $\Delta^{(AVA)}_{\rm pert}$ and
$\sigma^{(AVV)}_{\rm pert}$ of the perturbative three-point
graphs; as soon as the effective thresholds $s_{\rm eff}(Q^2)$ and
$\bar s_{\rm eff}(Q^2)$ have been fixed, extraction of the form
factors is straightforward. Formulating reliable criteria for
fixing the thresholds is, however, a very~difficult task,
discussed in great detail in \cite{lms}. For $Q^2\to\infty,$ the
form factors satisfy the factorization theorems\begin{eqnarray}
Q^2\,F_{\pi}(Q^2)\to8\pi\,\alpha_{\rm s}(Q^2)\,f_\pi^2\ ,\qquad
Q^2\,F_{\pi\gamma}(Q^2)\to\sqrt{2}\,f_\pi\ ,\qquad f_\pi=130\;
\mbox{MeV}\ .\end{eqnarray}Owing to some properties of the
spectral densities, this behaviour is correctly reproduced by
(\ref{Fld})~if\begin{eqnarray}\label{ass}s_{\rm
eff}(Q^2\to\infty)=\bar s_{\rm
eff}(Q^2\to\infty)={4\pi^2\,f_\pi^2}\ .\end{eqnarray}For finite
$Q^2$, however, the effective thresholds $s_{\rm eff}$ and $\bar
s_{\rm eff}$ depend on $Q^2$ and differ from each~other
\cite{lms}; the ``conventional LD model'' assumes (\ref{ass}) to
hold even down to values of $Q^2$ not too small~\cite{ld}.

\newpage\noindent Needless to say, such conventional LD model for
effective thresholds is an approximation not taking into account
details of the confining dynamics. Its only relevant feature is
factorization of hard form factors. Thus, it can be checked in
quantum mechanics, using potentials of Coulomb-plus-confining
shape for the pion's elastic form factor and of purely confining
shape for its transition form factor.

\section{Exact vs.\ Local-Duality Form Factors in
Quantum-Mechanical Potential Models}
\begin{figure}[!h]\begin{center}\begin{tabular}{cc}
\includegraphics[width=7.147cm]{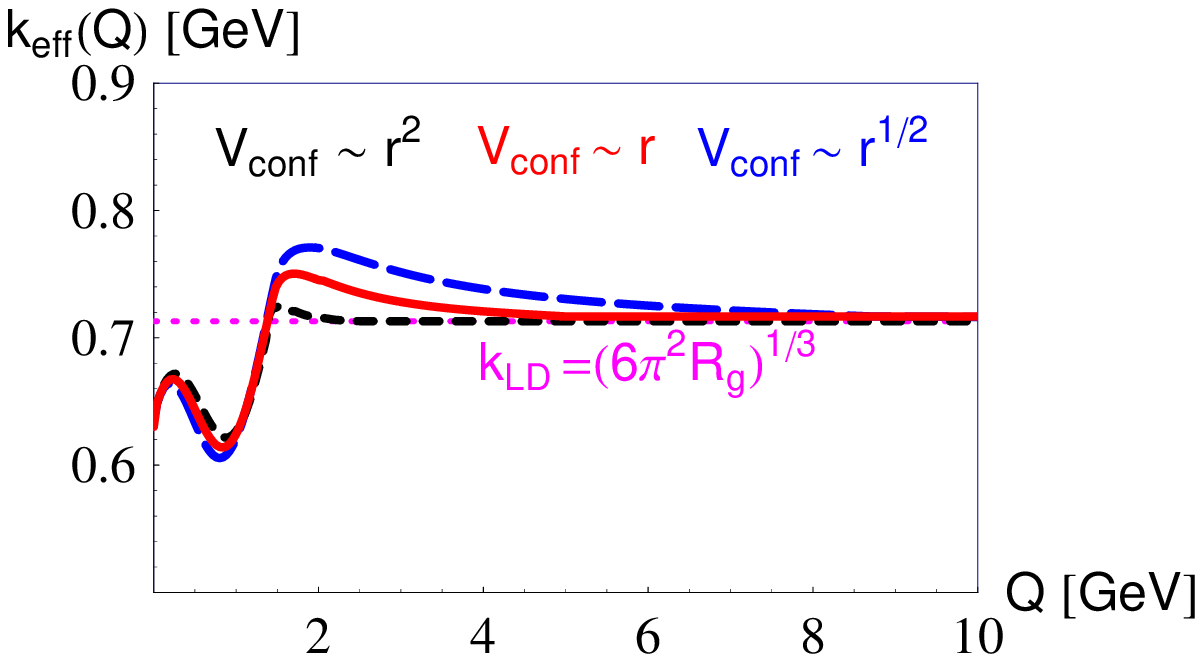}&
\includegraphics[width=7.147cm]{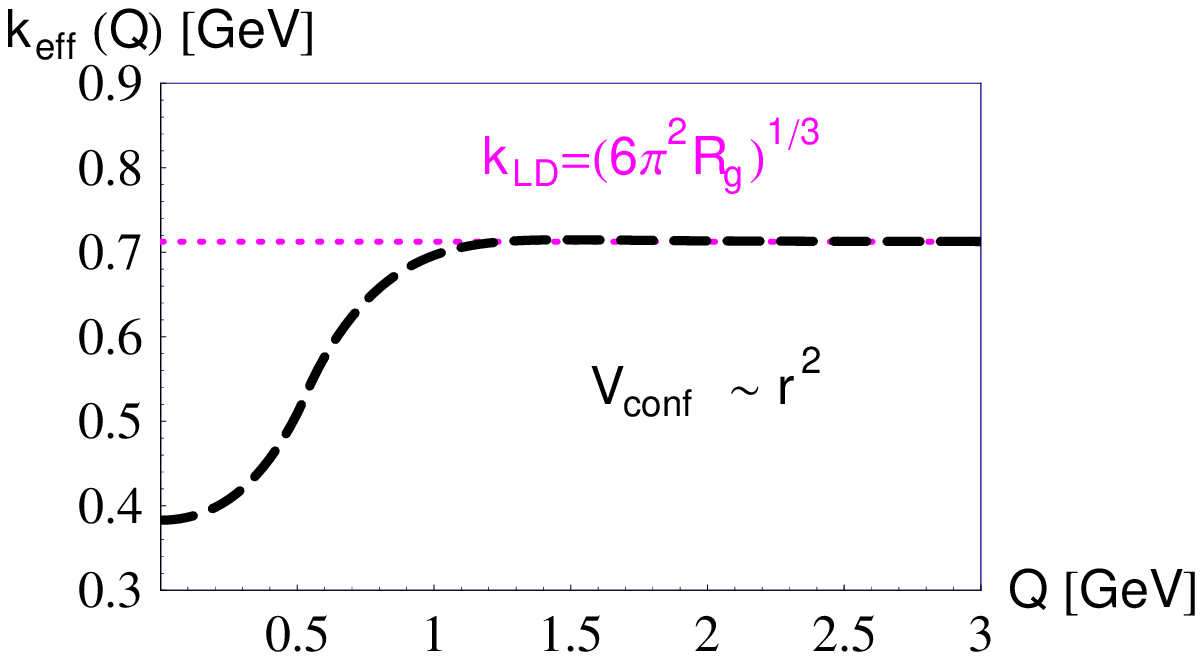}\end{tabular}
\caption{\label{Plot:2}QM exact effective thresholds for elastic
(left) and transition (right) form factors for different~$V_{\rm
conf}.$}\end{center}\end{figure}

Quantum-mechanical (QM)
potential models provide a possibility to test the accuracy
of~an~LD model by comparing the exact form factors, obtained from
the solution of the Schr\"odinger equation, with the outcomes of
this QM LD model constructed in precisely the same way as in QCD.
Figure \ref{Plot:2} shows the exact effective thresholds $k_{\rm
eff}$ that reproduce the exact form factors via the~LD~expression.
Irrespective of the confining interaction $V_{\rm conf}(r),$ the
precision both of the LD approximation~for the effective threshold
and of the LD elastic form factor increases with $Q^2$ in the
region $Q^2\ge5$--$8$~GeV$^2;$ for the transition form factor, the
LD approximation starts to work well at~even~smaller~values
of~$Q^2.$

\section{The Pion Elastic Form Factor $F_{\pi}(Q^2)$
\cite{blm2011}}

\begin{figure}[!b]\begin{center}\begin{tabular}{cc}
\includegraphics[width=7.147cm]{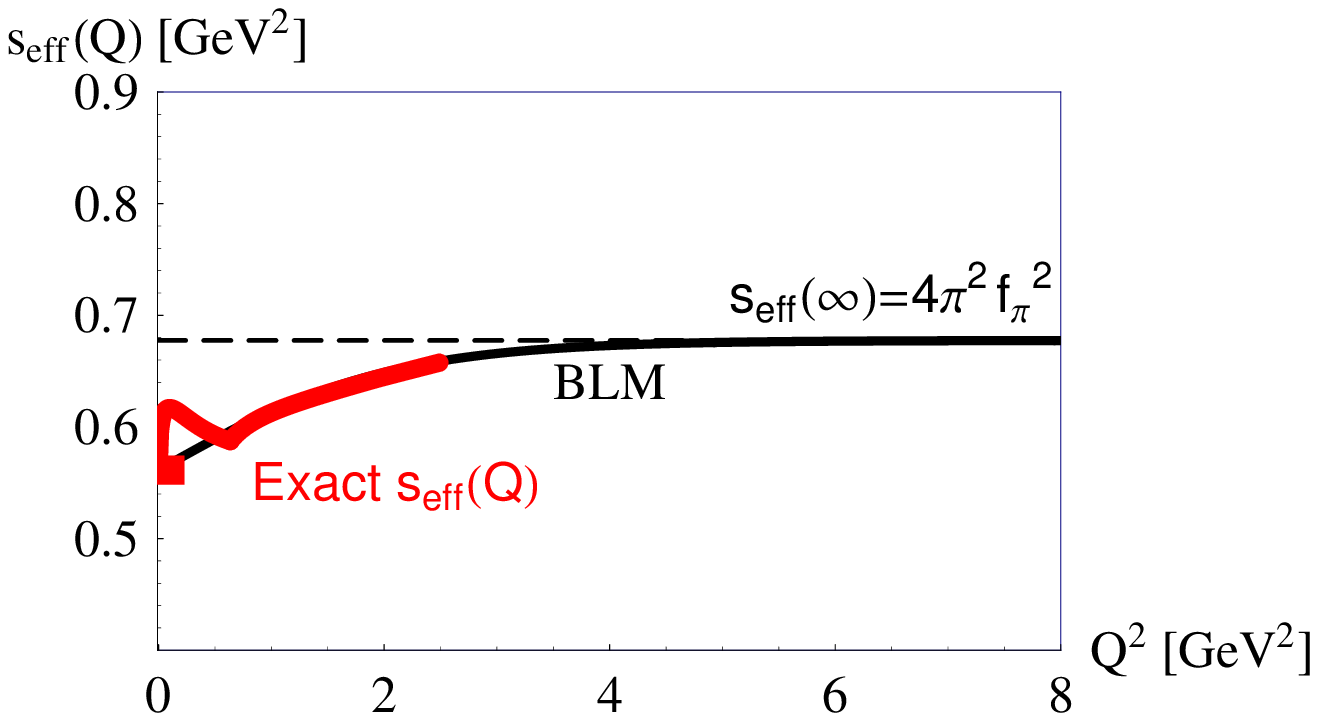}&
\includegraphics[width=7.147cm]{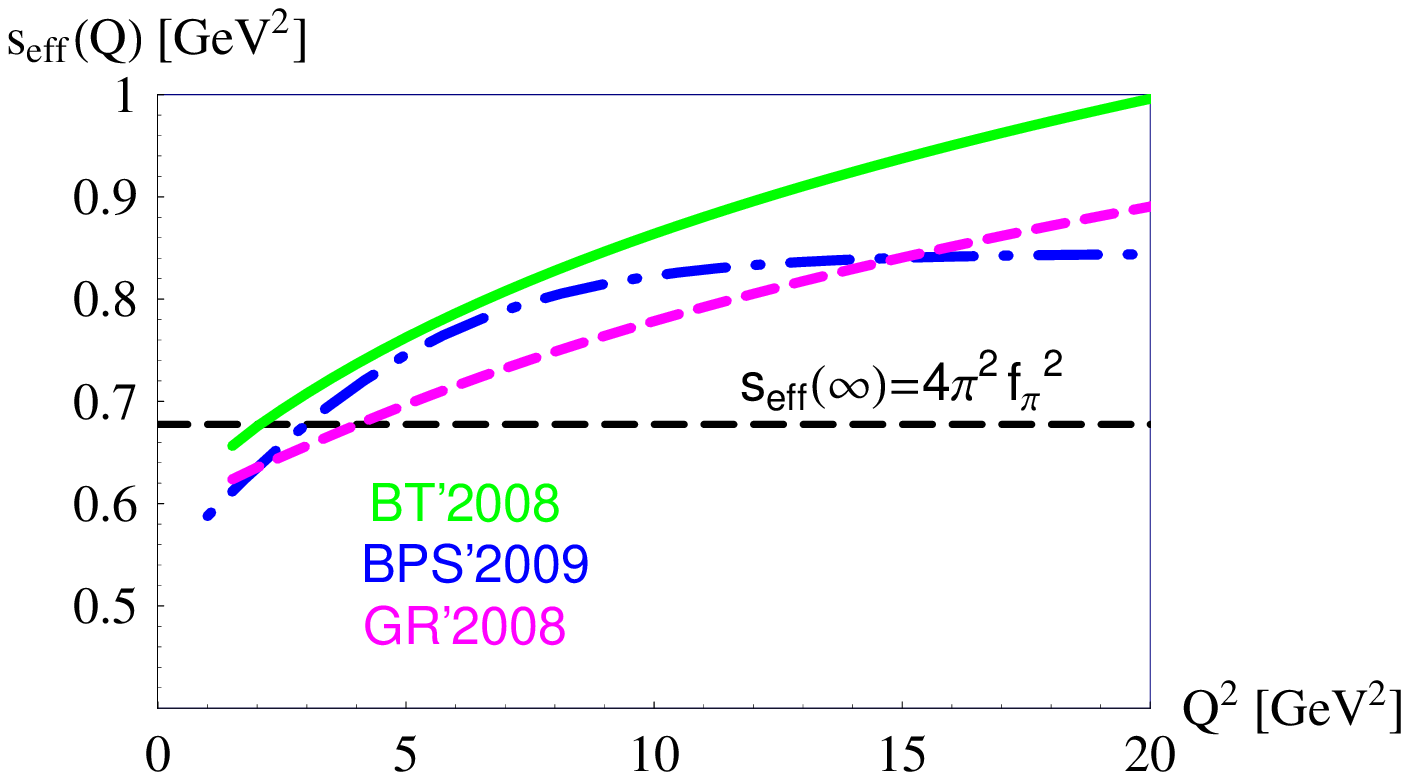}
\end{tabular}\caption{\label{Plot:3}Equivalent effective thresholds
$s_{\rm eff}$ for the pion elastic form factor extracted from the
experimental data \cite{data_piff} vs.\ the improved LD model of
\cite{blm2011} (left) and from the theoretical predictions
depicted in Fig.~\protect\ref{Plot:1}~(right).}\end{center}
\end{figure}

Let us introduce the notion of an {\em equivalent effective
threshold\/}, defined as that quantity $s_{\rm eff}(Q^2)$ which
reproduces by Eq.~(\protect\ref{Fld}) some preset behaviour of a
form factor. The exact effective threshold extracted from the data
(Fig.~\ref{Plot:3}) suggests that the LD limit might be reached
already at relatively low $Q^2,$ whereas its theoretical
counterparts imply that the accuracy of the LD model still
decreases with increasing $Q^2$ even at $Q^2$ as large as $Q^2=20$
GeV$^2,$ in conflict with our QM experience~and~the~hints from the
data at low $Q^2.$ Future more accurate JLab data in the range up
to $Q^2=8$ GeV$^2$ will decide.

\section{The $P\to\gamma\,\gamma^*$ ($P=\pi,\eta,\eta'$) Transition
Form Factors $F_{P\gamma}(Q^2)$ \cite{lm2011}}
\begin{figure}[!h]\begin{center}\begin{tabular}{cc}
\includegraphics[width=7.147cm]{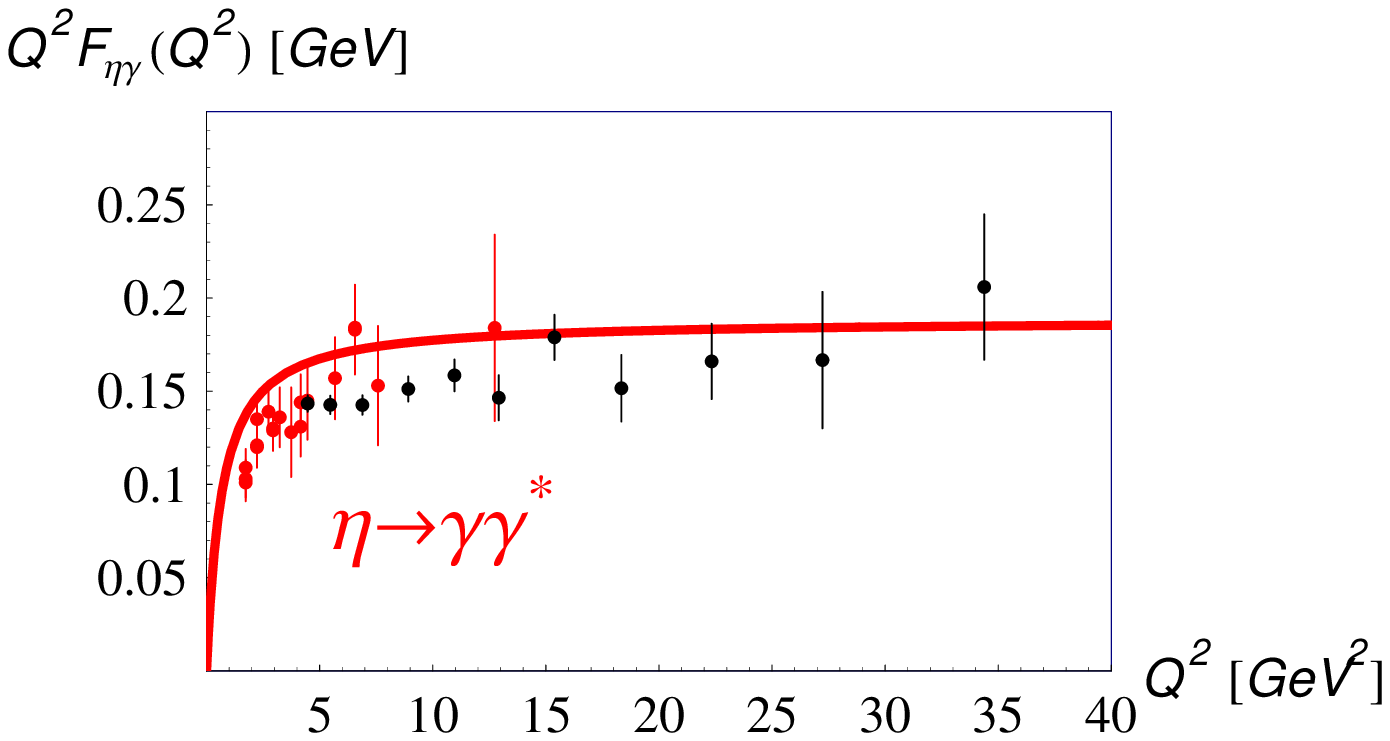}&
\includegraphics[width=7.147cm]{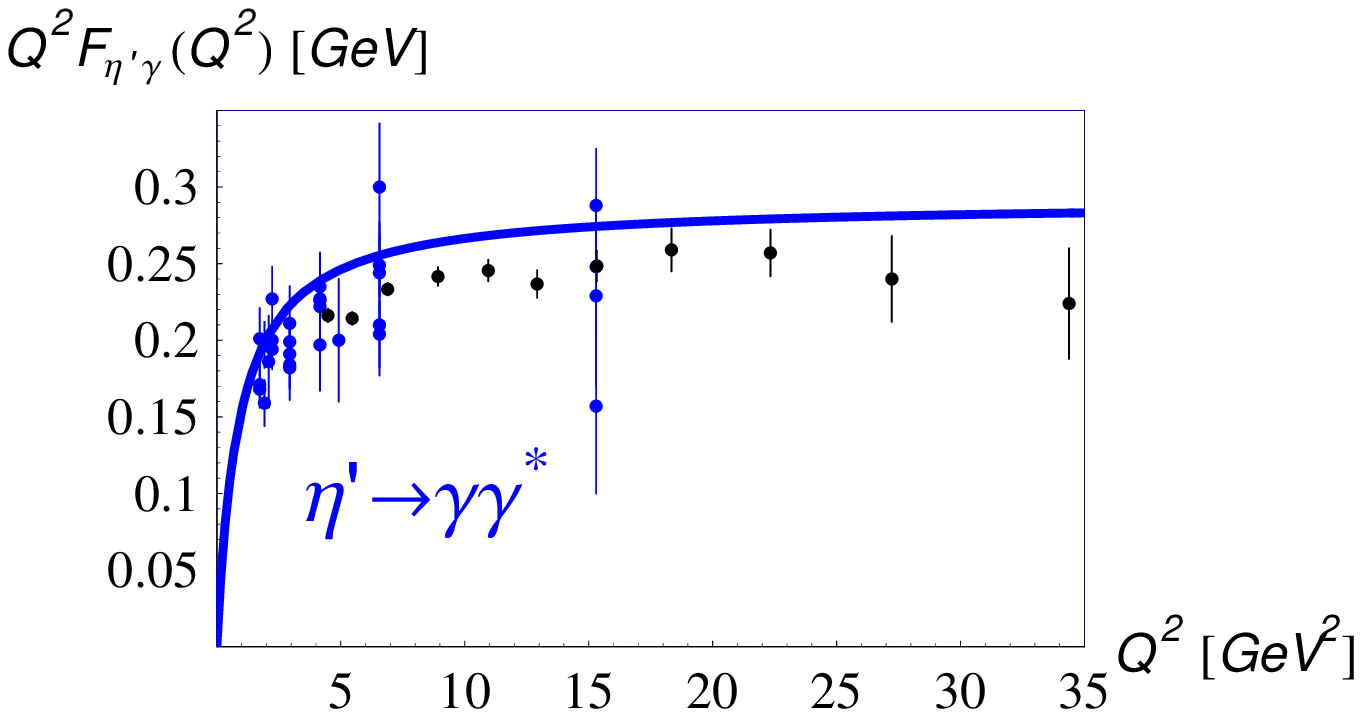}\end{tabular}
\caption{\label{Plot:4}LD predictions for both $\eta$ and $\eta'$
transition form factors $F_{(\eta,\eta')\gamma}(Q^2)$ vs.\
experimental data \cite{cello-cleo,babar1}.}\end{center}
\end{figure}

\begin{figure}[!b]\begin{center}\begin{tabular}{cc}
\includegraphics[width=7.147cm]{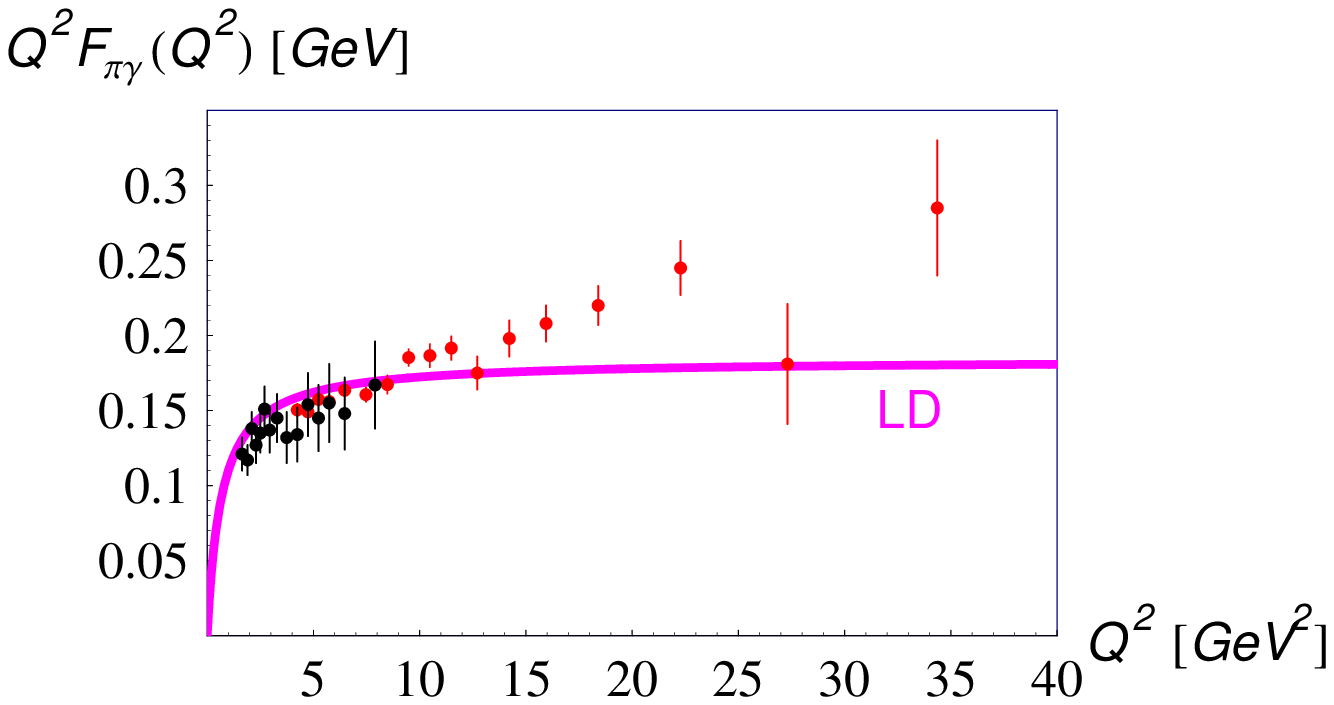}&
\includegraphics[width=7.147cm]{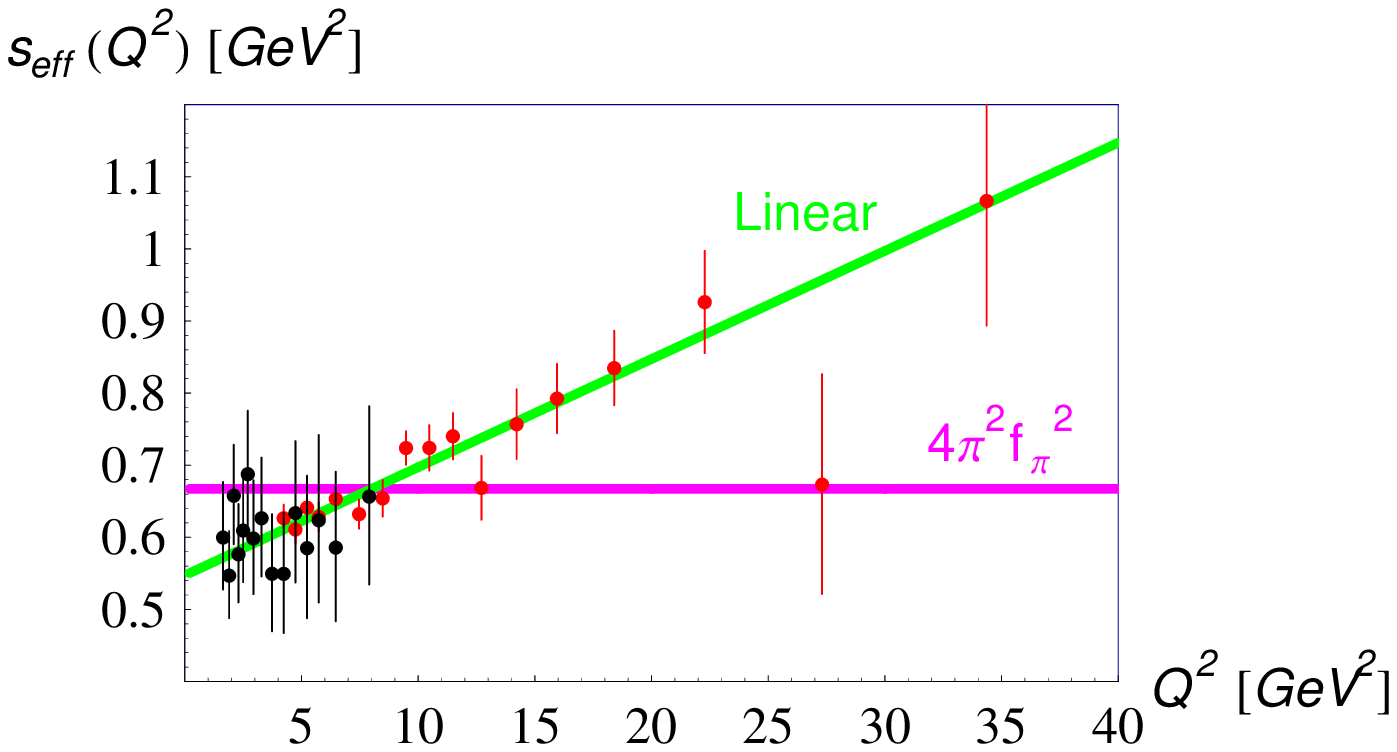}\end{tabular}
\caption{\label{Plot:5}$\pi\gamma$ transition form factor
$F_{\pi\gamma}(Q^2)$ vs.\ data \cite{cello-cleo,babar}, and
associated equivalent effective threshold $s_{\rm eff}.$}
\end{center}\end{figure}

For the $\eta$ and
$\eta'$ decays, we are obliged to take properly into account both
$\eta$--$\eta'$ mixing and the presence of two --- strange and
nonstrange --- LD form factors (for details, consult
\cite{anisovich,blm2011}). Figure~\ref{Plot:4} shows the
corresponding parameter-free predictions. There is an overall
agreement between the~LD model and the data. Surprisingly, for the
pion transition form factor (Fig.~\ref{Plot:5}) one observes a
manifest disagreement with the {\sc BaBar} data \cite{babar}.
Moreover, in distinct conflict with both the $\eta$ and $\eta'$
results and our QM experience, these data suggest that the LD
violations increase with $Q^2$ even in the range $Q^2\approx40$
GeV$^2$! It is hard to find a compelling argument explaining why
the nonstrange components in $\eta$ and $\eta',$ on the one hand,
and in $\pi^0$, on the other hand, should exhibit a such~different
behaviour.

\section{Summary and Conclusions}We reported the results of our
investigation of the pion elastic \cite{blm2011} and the
$\pi^0,\eta,\eta'$ transition~\cite{lm2011} form factors in the
framework of QCD sum rules in LD limit. Our main observations are
as follows:\begin{enumerate}\item For the elastic form factor, the
(approximate) LD model is expected to work increasingly well in
the region $Q^2\ge4$--$8$ GeV$^2$, independently of the details of
the confining interaction. For an arbitrary confining interaction,
this LD model reproduces the true form-factor behaviour very
precisely for $Q^2\ge20$--$30$~GeV$^2.$ Accurate data for the
pion's form factor indicate that the~LD value of its effective
threshold, $s_{\rm eff}(\infty)=4\pi^2\,f_\pi^2,$ is reached
already at relatively low~momenta $Q^2=5$--$6$ GeV$^2;$ rendering
large deviations from the LD limit for $Q^2=20$--$50$~GeV$^2$
unlikely.\item For all the $P\to\gamma\,\gamma^*$ transition form
factors, the LD model should work well for $Q^2$ larger~than a few
GeV$^2.$ Indeed, the LD model performs well for the
$\eta\to\gamma\,\gamma^*$ and $\eta'\to\gamma\,\gamma^*$ form
factors. For the $\pi\to \gamma\,\gamma^*$ form factor, however,
{\sc BaBar} data point to a {\em violation\/} of local duality,
rising with $Q^2,$ even at $Q^2$ as large as 40 GeV$^2,$
corresponding to an effective threshold of linear rise. So far,
this stunning puzzle withstood all attempts to find convincing
theoretical~explanations.\end{enumerate}

\vspace{4.74ex}\noindent{\bf Acknowledgments.} D.M.\ was supported
by the Austrian Science Fund (FWF), project no.~P22843.

\end{document}